\documentclass[twocolumn]{aastex631}

\shorttitle{Upper Limits to Proper Motions of an Orion JuMBO}
\shortauthors{Rodr\'\i guez et al.}

\graphicspath{{./}{figures/}}

\begin{document}

\title{Upper Limits to the Proper Motions of JuMBO 24, a Jupiter-Mass Binary Object Candidate in Orion}

\author[0000-0003-2737-5681]{Luis F. Rodr\'\i guez}
\affiliation{Instituto de Radioastronom\'\i{}a y
Astrof\'\i{}sica, Universidad Nacional Aut\'onoma de M\'exico \\
Apartado Postal 3--72, 58090 Morelia, Michoac\'an, M\'exico}
\affiliation{Mesoamerican Center for Theoretical Physics, 
Universidad Aut\'onoma de Chiapas \\
Carretera Emiliano Zapata Km 4, 29050 Tuxtla Guti\'errez, Chiapas, M\'exico}

\author[0000-0002-5635-3345]{Laurent Loinard}
\affiliation{Instituto de Radioastronom\'\i{}a y
Astrof\'\i{}sica, Universidad Nacional Aut\'onoma de M\'exico \\
Apartado Postal 3--72, 58090 Morelia, Michoac\'an, M\'exico}
\affiliation{Black Hole Initiative at Harvard University, 20 Garden Street, Cambridge, MA 02138, USA}
\affiliation{David Rockefeller Center for Latin American Studies, Harvard University,\\ 1730 Cambridge Street, Cambridge, MA 02138, USA}

\author[0000-0003-2343-7937]{Luis A. Zapata}
\affiliation{Instituto de Radioastronom\'\i{}a y
Astrof\'\i{}sica, Universidad Nacional Aut\'onoma de M\'exico \\
Apartado Postal 3--72, 58090 Morelia, Michoac\'an, M\'exico}

\author[0000-0002-2863-676X]{Gisela N. Ortiz-León}
\affiliation{Instituto Nacional de Astrofísica, Óptica y Electrónica, Apartado Postal 51 y 216, 72000 Puebla, Mexico}

\begin{abstract}
Using JWST near-infrared data of the inner Orion Nebula, \citet{PearsonMcCaughrean2023} detected 40 binary systems they proposed to be Jupiter-Mass Binary Objects (JuMBOs) -- although their actual nature is still in debate. Only one of the objects, JuMBO\,24, was detected in the radio continuum. Here, we report on new radio continuum (10 GHz) Karl G. Jansky Very Large Array (VLA) detections of the radio counterpart to JuMBO\,24, and on an unsuccessful search for 5 GHz continuum emission with the High Sensitivity Array (HSA). From our new VLA detections and adopting a distance to the region, we set an upper limit of $\simeq 6$~km~s$^{-1}$ to the velocity of the radio source in the plane of the sky. This upper limit favors an origin for this source similar to that of stars, that is, from a stationary contracting core. The nature of the radio emission remains uncertain but the lack of strong variability (all VLA observations are consistent with a steady flux of $\sim$50 $\mu$Jy), of detection on long HSA baseline, and of detectable circular polarization in VLA data do not favor a non-thermal origin. 
\end{abstract}


\section{Introduction} \label{sec:intro}

Using James Webb Space Telescope (JWST) near-infrared observations of the inner Orion Nebula and Trapezium Cluster, \citet{PearsonMcCaughrean2023} reported 40 binary systems that were proposed to be Jupiter-Mass Binary Objects (JuMBOs). These binary systems have separations in the plane of the sky between 28 and 384 au (0.07 and 0.99 arcsec at a distance of 388 pc; \citealt{Kounkel+2017}). The identification of these sources as having a planetary mass has been challenged by \citet{luh2024}, who proposes that they are actually reddened background stars. The resolution to this dilemma awaits for JWST spectroscopy.

\citet{rod2024} searched for radio continuum emission from the JuMBOs in Very Large Array observations and detected only one of them, object 24, in three epochs distributed over a decade, in 2012, 2018, and 2022. The analysis of these detections suggests that the radio flux of JuMBO\,24 is steady and that the emission comes from both components in roughly equal proportions because the peak in the radio map coincides, within the errors, with a central position between the two sources.

JuMBO\,24 was originally detected in near-infrared Hubble Space Telescope observations in the F160W (1.4-1.8 $\mu$m) and F110W (0.8-1.4 $\mu$m) filters, with magnitudes of 17.11 and 18.33, respectively \citep[][their source 274]{Luhman+2000}. \citet{Slesnick+2004} reported a K-band magnitude of 19.51 (their source 728). These authors used near-infrared photometry and spectroscopy as well as evolutionary stellar models \citep{Dantona_Mazzitelli_1994,Baraffe+1998} to determine an infrared spectral type of M5.5$\pm$1.5 and a mass of  $\sim 52$~M$_J$, under the assumption of a single star. The JWST data and the interpretation of \citet{PearsonMcCaughrean2023} would indicate, instead, that JuMBO\,24 is a binary with both components having the same giant-planet mass of $11.5~M_J$ and a separation of 28 au in the plane of the sky. JuMBO\,24 stands out in the list of these objects because it has the largest total mass ($23~M_J$) and the closest separation in the plane of the sky (28 au). Also, the primary of JuMBO\,24 has the second smallest extinction, $A_V$ = 3.6, behind only JuMBO\,27, that has $A_V$ = 2.4.

Here, we present new VLA observations of JuMBO\,24 taken in late 2024 and early 2025 as well as High Sensitivity Array (HSA) observations obtained in early 2024. This paper is organized as follows. In Section 2 we present the observations, while in Section 3 we discuss the implications of the new VLA detections and discuss the origin of the radio in light of the non-detection on HSA baselines. In Section 4 we summarize our conclusions.

\begin{deluxetable*}{cccccc}
\label{tab:obsvla}
\tablenum{1}
\tablecaption{Parameters of the VLA Observations\label{chartable}}
\tablewidth{900pt}
\tabletypesize{\scriptsize}
\tablehead{
\colhead{Mean} &  
\colhead{Synthesized Beam} & 
\colhead{Flux Density of} & \multicolumn{2}{c}{Position}  & Upper \\ 
\colhead{Epoch} & 
\colhead{($\theta_{max} \times \theta_{min}; PA$)} & \colhead{JuMBO~24 ($\mu$Jy)} & \colhead{RA(J2000)}  & 
\colhead{DEC(J2000)} & \colhead{Limit to $\vert V \vert$} } 
\decimalcolnumbers
\startdata
2024.806 &  
$0\rlap.{''}21 \times 0\rlap.{''}16;+30^\circ$ & 45.0$\pm$13.0 & $19\rlap.^s5068 \pm 0\rlap.^s0010$ & 
$39\rlap.{''}727 \pm 0\rlap.{''}009$ &  $\leq 70\%$ \\
2025.000 &  
$0\rlap.{''}19 \times 0\rlap.{''}15;+0^\circ$ & 65.3$\pm$8.7 & $19\rlap.^s5068 \pm 0\rlap.^s0006$ & 
$39\rlap.{''}720 \pm 0\rlap.{''}007$ &  $\leq 22\%$ \\
\enddata
\tablecomments{Observations are part of project 24B-082. They were made at 10 GHz with a bandwidth of 4 GHz (8 to 12 GHz). The position given in columns 
4 and 5 show only the seconds of the RA and the arcseconds of the DEC. The remaining values are
RA(J2000) = $05^h~19^m$ and DEC(J2000) = $-05^\circ~23'$. Column 6 gives the 4-$\sigma$ upper limit to the absolute 
circular polarization.
}
\end{deluxetable*}

\section{Observations} \label{sec:obs}
\subsection{VLA Observations and results}

The new VLA observations were made at $\nu = $ 10 GHz as part of project 24B-082 on six epochs: 2024 October 21, December 21 and 28 (two observations in this epoch), and 2025 January 7 and 9. All observations were made in the highest angular resolution A configuration. The last five epochs were taken over a period of only 20 days and were concatenated to produce a single, high signal-to-noise, data point. The absolute flux calibrator was J0137$+$331 (3C\,48) and the gain calibrator was J0541$-$0541. The parameters of these observations are summarized in Table \ref{tab:obsvla}. The data were calibrated in the standard manner using the CASA (Common Astronomy Software Applications; \citealt{McMullin+2007}) package of NRAO and the pipeline provided for VLA\footnote{https://science.nrao.edu/facilities/vla/data-processing/pipeline} observations. A robust weighting of 0 \citep{Briggs_1995} was used in the imaging to optimize the compromise between sensitivity and angular resolution. The images were made with a $(u,v)$ range set to $>100~k\lambda$, to suppress structures larger than 2$''$ that limit the signal-to-noise ratio in the Orion field. We detected JuMBO\,24 in all six epochs with a flux density consistent with $\sim$50 $\mu$Jy. An image made of the concatenation of the last five epochs (that were observed within a period of 19 days) is shown in Figure \ref{fig:VLAmap}. In this Figure the cross indicates the position of JuMBO 24 precessed to the same average epoch of the radio image. \citet{rod2024} concluded that the \sl a priori \rm probability of one Orion JuMBO randomly coinciding in position with one Orion radio source was only 0.00012, supporting a real association between JuMBO\,24 and the radio source. These authors also show an overlap of the 2024 radio image and the JWST image.

\subsection{HSA Observations}\label{sec:HSA}
JuMBO\,24 was also observed as part of project BL314 on 22 January 2024 at $\nu = 5$ GHz with the High Sensitivity Array (HSA) which combines the VLBA antennas with the phased VLA array and the Green Bank Telescope. The data were recorded at a rate of 4 Gbps corresponding to an aggregate bandwidth of 1 GHz. The gain calibrator was J0541$-$0541. The data were calibrated and imaged in AIPS following standard procedures. The synthesized beam has a size of 13 $\times$ 2 mas at PA = --8.8$^{\circ}$. The final image has a noise level of 6 $\mu$Jy beam$^{-1}$. No source was detected, setting a 3-$\sigma$ upper limit on the flux density recovered on HSA baselines of 18 $\mu$Jy. 

\section{Discussion} \label{sec:discussion}

\subsection{Upper Limits to the Proper Motion}

\begin{figure*}
\plottwo{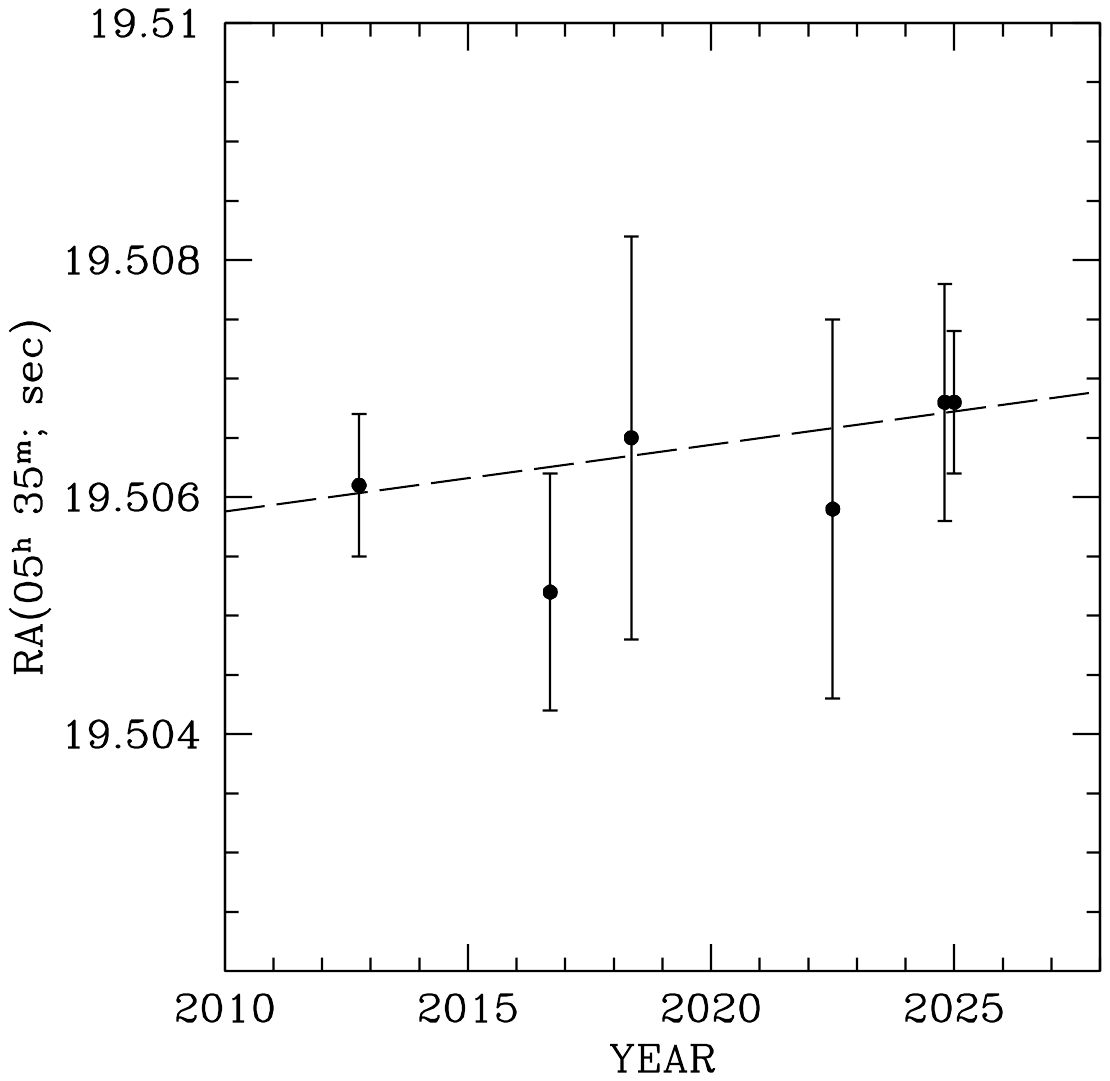}{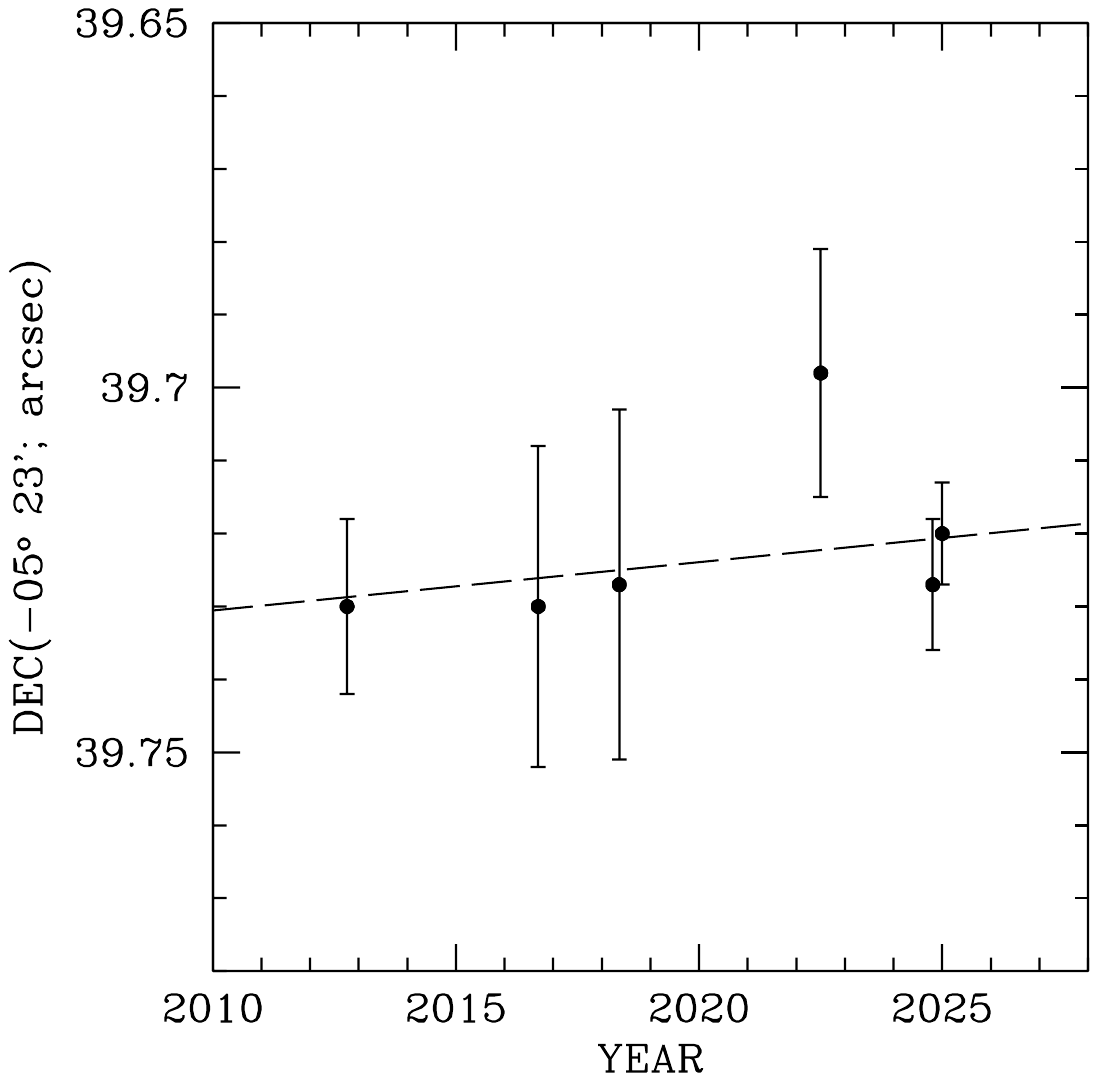}
\vskip-2.0cm
\caption{Radio position of JuMBO\,24 as a function of time. The dashed lines indicate the least-squares linear fit to the data.}
\label{fig:VLApm}
\end{figure*}

As the first radio continuum detection of a possible planetary-mass binary object, the study of JuMBO\,24 is relevant. Among the results of \citet{rod2024} an upper limit to the proper motions of the source was obtained. We combined the four data points of \citet{rod2024} with the two new data points given in Table \ref{tab:obsvla} to search for proper motions in the source (Figure \ref{fig:VLApm}). From these data we obtain more stringent upper limits:

$$\mu_{\alpha} \cos \delta = +1.01 \pm 0.94 \text{~mas~yr}^{-1} ~~;$$
$$ ~~\mu_\delta  = +0.66 \pm 1.03\text{~mas~yr}^{-1}.$$

\noindent
This result implies that there are no significant proper motions at a 3-$\sigma$ level of $\sim$3 mas~yr$^{-1}$, that corresponds to a transverse velocity of $\sim 6$ km~s$^{-1}$ at a distance of 388 pc \citep{Kounkel+2017}. These upper limits rule out large velocities in the plane of the sky, although it should be pointed out that significant velocities could be present in the radial direction. Our measurements can be registered to an approximate rest frame for the Orion Cluster by subtracting the average proper motion of the radio stars in the core of Orion \citep{Dzib+2017}, $\mu_{\alpha} \cos\delta = +1.07 \pm 0.09\text{~mas~yr}^{-1}; \mu_{\delta}  = -0.84 \pm 0.16\text{~mas~yr}^{-1}$. We obtain:

$$\mu_{\alpha} \cos \delta = -0.06 \pm 0.94\text{~mas~yr}^{-1} ~~ ;$$
$$ ~~ \mu_\delta  = +1.50 \pm 1.04\text{~mas~yr}^{-1}.$$

\begin{figure*}
\epsscale{1.0}
\vspace{-6cm}
\plotone{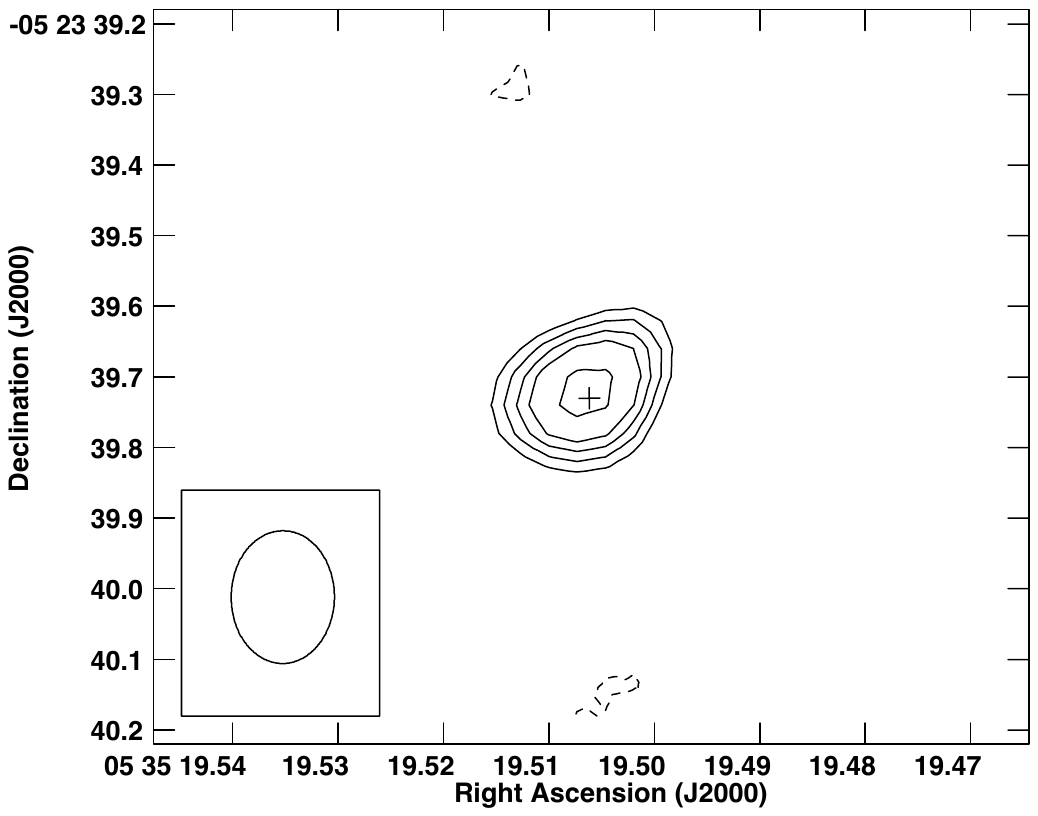}
\caption{ \vspace{-0cm} VLA image of the 10 GHz emission from JuMBO\,24, made concatenating the last five epochs of project 24B-082. The contours are -3, 3, 4, 5, 6, and 8 times 4 $\mu$Jy~beam$^{-1}$, the rms of the image. The cross indicates the position of JuMBO\,24 as reported by \citet{PearsonMcCaughrean2023}: RA(J2000) = $05^h 35^m 19\rlap.^s50616$, DEC(J2000) = $-05^\circ 23' 39\rlap.{''}7303$. The synthesized beam of the radio map is shown at the bottom left of the image.}
\label{fig:VLAmap}
\end{figure*}

\noindent
We conclude that JuMBO\,24 is not moving at large velocities compared to the radio stars in Orion. \citet{2019A&A...624A.120V} have made numerical simulations for the evolution of young and dense stellar clusters.They find that  single free-floating planets ejected from their original system after a strong encounter with another star will have velocities typically three times those of the stars in the cluster. \citet{Dzib+2017} find a typical velocity dispersion of $\sim 2-3$ km~s$^{-1}$ along a plane-of-the-sky coordinate for the radio-emitting stars in the core of the Orion Nebula Cluster. Assuming that the results of \citet{2019A&A...624A.120V} can be extended to free-floating binary planets, we expect velocity dispersions of $\sim 6-9$ km~s$^{-1}$ for the JuMBOs as well as for the $\sim$500 single planetary-mass objects reported in Orion by \citet{PearsonMcCaughrean2023}. Our upper limits favor an origin for the JuMBOs similar to that of stars, that is, from the contraction of a stationary core. \citet{2024dia} have proposed that JuMBOs could be formed from the photoerosion of prestellar cores by Lyman continuum radiation from nearby massive stars. In this model the JuMBOs would be stationary with respect to their environment. Future highly accurate determinations of the kinematics of the Orion JuMBOs will help understand the origin of these intriguing sources.

\subsection{Morphology of the Radio Source}

\citet{rod2024} performed a Gaussian ellipsoid fit to their 6 GHz radio image using the task IMFIT of CASA to obtain a deconvolved angular size of $0\rlap.{''}179 \pm 0\rlap.{''}045 \times 0\rlap.{''}143  \pm 0\rlap.{''}074; 60^\circ \pm 89^\circ$. The same procedure applied to our higher angular resolution 10 GHz image (Figure \ref{fig:VLAmap}) gives a deconvolved angular size of $0\rlap.{''}176 \pm 0\rlap.{''}033 \times 0\rlap.{''}039  \pm 0\rlap.{''}062; 108^\circ \pm 15^\circ$. These angular dimensions and the orientation of the major axis are consistent with the value obtained from Gaussian ellipsoid fits to the JWST images. For instance, a fit to the image at 2.77 $\mu$m results in $0\rlap.{''}170 \pm 0\rlap.{''}002 \times 0\rlap.{''}132  \pm 0\rlap.{''}002; 97^\circ \pm 2^\circ$. The shorter wavelength JWST observations show that the cause of the extension along the east-west direction is the binarity of the source, which is comprised of two objects separated by about 100 mas closely along the east-west direction. This result suggests that, as in the near-infrared, there is comparable radio emission from both components of the binary. This result is similar to that obtained for the compact ultra-cool dwarf binary VHS 1256$-$1257AB, where the optical and radio emissions show compatible dimensions and orientation \citep{Rodriguez+2023}.

\subsection{Nature of the radio emission}

The origin of the radio emission in JuMBO\,24 remains uncertain. The fairly flat spectral index between 6 and 10 GHz documented by \citet{rod2024} is consistent with both thermal free-free emission and non-thermal (e.g.\ gyro-synchrotron) radiation. Two direct ways to ascertain a non-thermal origin would be (i) detections on very long base baselines ($>$ 5,000 km), which would require brightness temperatures in excess of 10$^6$ K, and (ii) detection of polarization. As mentioned in Section \ref{sec:HSA}, we failed to detect emission in our HSA observations, setting a 3$\sigma$ upper limits of about 18 $\mu$Jy. This clearly rules out the existence of a 50 $\mu$Jy single source but cannot entirely exclude the possibility of two 25 $\mu$Jy sources. Regarding circular polarization, none was detected at the level of about 20\% in our deep 10 GHz integration obtained by concatenating all VLA data obtained in early 2025 (Table \ref{tab:obsvla}). We note, finally, that another frequent properties on non-thermal radiation, strong variability associated with flares, is not seen in JuMBO\,24 either since all VLA detections are consistent with a constant flux density of $\sim 50$ $\mu$Jy. Although no final conclusion can be drawn from these results regarding the nature of the radio emission, the properties reported here at least do not favor a non-thermal origin.

\section{Conclusions}

We report new radio continuum VLA detections at 10 GHz toward the Jupiter-Mass Binary Object candidate JuMBO\,24 in the Orion Nebula Cluster. Within the errors, the radio emission appears steady on timescales from days to years and exhibits no large circular polarization. No detection was achieved in HSA observations, ruling out the possibility that JuMBO\,24 is a single 50 $\mu$Jy non-thermal source, but not conclusively that it could be a  binary system comprised of two 25~$\mu$Jy non-thermal objects. No significant proper motion is detected between the four observing epochs, ruling out large velocities ($\geq$ 6~km~s$^{-1}$) in the plane of the sky. The radio emission is marginally resolved in the same direction as the infrared source detected by the JWST, suggesting that the radio emission comes from similar contributions of the two planetary mass candidate objects. Additional radio observations are necessary to pin down the nature of the radio emission mechanism and to further restrict its proper motion.

\bibliography{JuMBO25}{}
\bibliographystyle{aasjournal}

\begin{acknowledgments}
The National Radio Astronomy Observatory is a facility of the National Science Foundation operated under cooperative agreement by Associated Universities, Inc. We acknowledge the support from DGAPA, UNAM (projects IN112323, IN112417, IN112820 and IN108324) and CONAHCyT, México (projects 238631, 280775 and 263356).
\end{acknowledgments}

\vspace{5mm}
\facilities{NASA/ESA/CSA(JWST), NRAO(VLA)}

\software{AIPS \citep{Greisen_2003}, CASA \citep{McMullin+2007}}

\end{document}